\documentclass[twocolumn,showpacs,preprintnumbers,amsmath,amssymb,groupedaddress,prl]{revtex4}

\usepackage{graphicx}
\usepackage{dcolumn}
\usepackage{bm}
\usepackage{epsfig}

\def\e{\begin{equation}}
\def\f{\end{equation}}
\def\=#1{\overline{\overline #1}}

\def\-#1{{\bf #1}}
\def\.{\cdot}
\def\l#1{\label{eq:#1}}
\def\r#1{(\ref{eq:#1})}

\begin{document}

\title{Enhancement of evanescent waves inside media with extreme optical anisotropy}

\author{Pavel A. Belov}
\affiliation{Department of Electronic Engineering, Queen Mary
University of London, Mile End Road, London, E1 4NS, United Kingdom}

\author{Yan Zhao}
\affiliation{Department of Electronic Engineering, Queen Mary
University of London, Mile End Road, London, E1 4NS, United Kingdom}

\author{Yang Hao}
\affiliation{Department of Electronic Engineering, Queen Mary
University of London, Mile End Road, London, E1 4NS, United Kingdom}

\author{Clive Parini}
\affiliation{Department of Electronic Engineering, Queen Mary
University of London, Mile End Road, London, E1 4NS, United Kingdom}

\begin{abstract}
Significant enhancement of evanescent spatial harmonics inside the
slabs of media with extreme optical anisotropy is revealed. This
phenomenon results from the pumping of standing waves and has the
feature of being weakly sensitive to the material losses. Such
characteristics may enable subwavelength imaging at considerable
distances away from the objects.
\end{abstract}

 \maketitle

Modern nanofabrication techniques have brought major breakthroughs
in the area of photonics and nanotechnology and make it possible to
manufacture structures with details less than a few nanometers. Such
advances in manufacturing technology lead to further miniaturization
of optical systems and stimulate increasing interest in the
development of devices capable of manipulating electromagnetic waves
at the subwavelength scale: e.g. optical waveguides with their
cross-sections much smaller than the wavelength, structures enabling
nanoscale concentration of light, superlenses with resolutions below
the diffraction limit, etc. All these devices are capable of
manipulating the near fields and they dramatically differ from
conventional optical components that mostly deal with the
propagating waves. In this letter we report a novel means to enhance
evanescent waves which can overcome a major constraint of optical
near field devices suffering from the rapid spatial decay of
evanescent harmonics. One possible application is the performance
improvement of near-field scanning optical microscopy. Currently,
this technique suffers from the limitation that the objects of
interest have to be located close enough to the probe of the
scanner, otherwise, the near field of the object which contains
subwavelength details will be too weak to be detected. The use of
the proposed evanescent wave enhancement effect will allow the
near-field scanners to operate with the objects located at
significant distances. This important advance opens the door to a
variety of possible applications, including the case when physical
access to the near field region of an object is not feasible. For
example, sub-surface capillary vessels in biological systems can
potentially be accessed using non-pervasive methods by exploiting
the near-field of light.

It is known that evanescent waves can be enhanced by the use of
resonant excitation of surface plasmons. Thin slabs of noble metals
can ``amplify'' evanescent harmonics within frequency ranges of
their plasmonic resonances \cite{Silversub}. Theoretically,
significant ``amplification'' can be provided by slabs of
metamaterials with negative permittivity and permeability, also
known as perfect lenses \cite{Pendrylens}. However, the practical
application of the perfect lens concept has been stalled by the
limitations of loss and bandwidth \cite{Narimanov} and the creation
of metamaterials in the optical domain remains a major challenge
\cite{Shalaev}. In this letter, we propose a conceptually different
approach to enable the significant enhancement of evanescent
harmonics. Its principle is based on pumping of standing waves
inside a slab of extremely anisotropic medium and features weak
sensitivity to material losses.

Let us consider an uniaxial dielectric medium with permittivity
dyadic of the form:
 \e \overline{\overline \varepsilon }  =\left(
               \begin{array}{ccc}
                  \varepsilon_{xx}& 0 & 0 \\
                 0 & \varepsilon & 0 \\
                 0 & 0 & \varepsilon \\
               \end{array}
             \right). \l{eps_aniso} \f
The dispersion equation for extraordinary waves in such medium
reads:
 \e \frac{k_x^2}{\varepsilon}+\frac{k_y^2+k_z^2}{\varepsilon_{xx}}=\frac{\omega^2}{c^2}, \l{disper} \f
where $k_x,k_y,k_z$ are the components of wave vector $\- k$,
$\omega$ is the frequency of operation and $c$ is the speed of light
in the vacuum.

If the medium has extreme optical anisotropy, that means
$|\epsilon_{xx}|\gg \epsilon$, then \r{disper} has the following
solution: \e k_x=\pm \sqrt{\varepsilon}\frac{\omega}{c} \mbox{ and }
k_y,k_z \mbox{ are arbitrary}. \l{dispersionless}\f Therefore, all
extraordinary waves of the medium with extreme optical anisotropy
are dispersionless: they travel ultimately along the direction of
anisotropy with a fixed phase velocity irrespectively of their
transverse variation. Note, that $\epsilon_{xx}$ must not
necessarily be real. It can be a complex number with a large
absolute value. The imaginary part of $\epsilon_{xx}$ representing
losses has no influence on the properties of waves in the medium.

Basically, a whole class of dielectrics with extreme optical
anisotropy can be accurately described by permittivity dyadic in the
form
 \e \overline{\overline \varepsilon }  =\left(
               \begin{array}{ccc}
                  \infty& 0 & 0 \\
                 0 & \varepsilon & 0 \\
                 0 & 0 & \varepsilon \\
               \end{array}
             \right). \l{eps_infty} \f
The infinite value appearing in \r{eps_infty} should not be treated
as something exceptional or unusual. The corresponding material
relations read as $E_x=0$, $D_{y}=\varepsilon E_{y}$ and
$D_{z}=\varepsilon E_{z}$. This actually means that the medium is
perfectly conducting along $x$-direction.

The latter fact gives a hint how media with extreme optical
anisotropy can be created artificially. An array of parallel
metallic wires, also known as a wire medium \cite{WMPRB}, will
provide necessary properties at frequencies up to the infrared band
\cite{Silv_Nonlocalrods}. In the visible frequency range, one of the
options may be a layered-metal dielectric structure
\cite{layeredPRB} which has an effective permittivity in the form
\r{eps_infty}.

Let a slab with thickness $d$ of medium with extreme optical
anisotropy in a direction perpendicular to the interfaces be excited
by an obliquely incident $p$-polarized plane wave with the
tangential component of the wave vector $k_t$. The time dependence
is $e^{-i\omega t}$. The total magnetic field can be written in the
form:
 \e H(x)=H_{i}\left\{
\begin{array}{lcl}
e^{i\sqrt{k^2-k_t^2}x} + R e^{-i\sqrt{k^2-k_t^2}x},\quad x<0 \\[2mm]
\begin{array}{lcl}
Ae^{i\sqrt{\varepsilon}kx}+ Be^{-i\sqrt{\varepsilon}kx},\\
\end{array} \quad 0\le x\le d \\[2mm]
T e^{i\sqrt{k^2-k_t^2}(x-d)},\quad x>d, \vphantom{\frac{A}{A}}\\
\end{array}\right.
\l{H} \f where $x$ is the coordinate along anisotropy axis, $k$ is
the wave number of free space, $H_i$ is the amplitude of the
incident wave, $R$ and $T$ are the reflection and transmission
coefficients, $A$ and $B$ are the amplitudes of waves inside of the
slab traveling in forward and backward directions, respectively. The
values of these coefficients can be easily obtained by matching
boundary conditions at the interfaces:
\begin{eqnarray}
  A &=& 2\sqrt{1-\frac{k_t^2}{k^2}}\left(\frac{1}{\sqrt{\varepsilon}}+\sqrt{1-\frac{k_t^2}{k^2}}\right)/D \nonumber \\
  B &=& 2\sqrt{1-\frac{k_t^2}{k^2}}\left(\frac{1}{\sqrt{\varepsilon}}-\sqrt{1-\frac{k_t^2}{k^2}}\right)e^{-2j\sqrt{\varepsilon}
kd}/D \nonumber \\
  R &=& (e^{-2jkd}-1)\left(\frac{1}{\varepsilon}-\left[1-\frac{k_t^2}{k^2}\right]\right)
/D \nonumber\\
T &=&
4\sqrt{1-\frac{k_t^2}{k^2}}\frac{1}{\sqrt{\varepsilon}}e^{-j\sqrt{\varepsilon}
kd}/D, \nonumber \end{eqnarray} where
$$D=\left(\frac{1}{\sqrt{\varepsilon}}+\sqrt{1-\frac{k_t^2}{k^2}}\right)^2-\left(\frac{1}{\sqrt{\varepsilon}}-\sqrt{1-\frac{k_t^2}{k^2}}\right)^2
e^{2i\sqrt{\varepsilon} kd}.$$

If the thickness of the slab fulfills the Fabry-Perot resonance
condition $\sqrt{\varepsilon} kd=n\pi$ then
$$
  A =
  \frac{1}{2}\left(1+\sqrt{\varepsilon}\sqrt{1-\frac{k_t^2}{k^2}}\right),\
  B = \frac{1}{2}\left(1-\sqrt{\varepsilon}\sqrt{1-\frac{k_t^2}{k^2}}\right),$$
$$  R = 0, \qquad T = (-1)^n. $$

Note, that under this condition the slab operates in the
canalization regime \cite{canal,SWIWM,WMIR,layeredPRB} as a perfect
transmission device: it does not produce any reflections and
perfectly transmits arbitrary field distributions from the front- to
back-interface. This occurs since in the case of media with extreme
optical anisotropy the Fabry-Perot condition is verified
simultaneously for all spatial harmonics. This extraordinary
property is justified by the fact that all waves travel across the
slab with the fixed phase velocity irrespectively of their
transverse wave vectors \r{dispersionless}. Thus, the total
electrical length of the slab remains the same for all possible
angles of incidence.

The effect is observed in the slabs with their thicknesses equal to
an integer number of half-wavelengths. In the next part of the
letter, for simplicity, we only consider slabs of a half-wavelength
thickness: $d=\lambda/(2\sqrt{\varepsilon})$.

The magnetic field distribution inside the slab can be written in
the following form \e H(x)=H_{i}
\left[\cos(\sqrt{\varepsilon}kx)+i\sqrt{\varepsilon}\sqrt{1-\frac{k_t^2}{k^2}}\sin(\sqrt{\varepsilon}kx)\right].
\l{Hx}\f For the evanescent part of spatial spectrum ($|k_t|>k$)
this distribution has the form of a standing wave with
amplitude\vspace{-2mm}\e H_m=H_i
\sqrt{1+\varepsilon\left(\frac{k_t^2}{k^2}-1\right)}.\f

\begin{figure}[htb] \centering
\includegraphics[width=8cm]{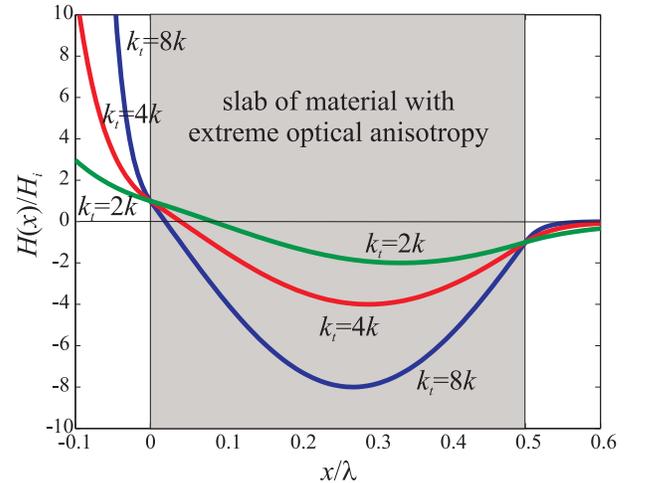}
\caption{(Color online) Distributions of magnetic field around and
inside of a slab of medium with extreme optical anisotropy excited
by evanescent waves with various values of the tangential component
of the wave vector $k_t$.} \label{enhance}
\end{figure}

This amplitude is a growing function of $k_t$. For high spatial
harmonics ($k_t\gg k$) it has a nearly linear dependence $H_m\approx
\sqrt{\varepsilon} k_t/k$. This means the higher $k_t$ is, the more
rapidly the evanescent wave decays and the higher amplitude of
standing wave is excited inside the slab. This fact is illustrated
in Fig. \ref{enhance}.

Effectively, this phenomenon can be treated as an increase in the
amplitude of the evanescent wave. Such an enhancement is observed
only inside the slab, but not outside as in alternative
configurations with slabs of materials with negative material
parameters \cite{Silversub,Pendrylens}.
\begin{figure}[htb] \centering
\includegraphics[width=8cm]{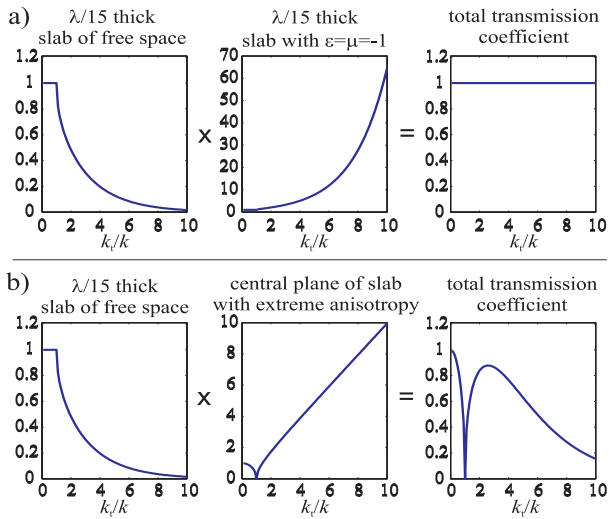}
\caption{(Color online) Comparison of absolute values of
transmission characteristics of slabs formed (a) by metamaterial
with $\varepsilon=\mu=-1$ ($\lambda/15$ thick) and (b) by medium
with extreme optical anisotropy as functions of transverse wave
vector $k_t$ and their efficiency to compensate decay of evanescent
waves in a $\lambda/15$ thick slab of free space. } \label{compare}
\end{figure}
In order to further compare the enhancement capabilities of these
two alternative approaches, the transmission coefficients for slabs
formed by a medium with $\varepsilon=\mu=-1$ and by a medium with
extreme optical anisotropy are presented in Fig. \ref{compare} as
functions of transverse wave vector. In the case of metamaterial
slab, the coefficient describes the transmission from the front- to
back-interface. Whereas, for the slab of medium with extreme optical
anisotropy the transmission coefficient is calculated from the
front-interface to the central plane. According to \r{Hx} it is
equal to: \e T(d/2)=\sqrt{\varepsilon}\sqrt{\frac{k_t^2}{k^2}-1}.
\l{Td2}\f The growing exponential and nearly linear dependencies for
transmission coefficients under comparison are clearly visible in
Fig. \ref{compare}. The slab of metamaterial is chosen to have a
thickness equal to $\lambda/15$ which is able to completely
compensate for the decay of evanescent waves in free space at a
$\lambda/15$ distance. In the case of slab formed by medium with
extreme optical anisotropy, the total transmission coefficient does
not equal to unity for any transverse wave vector in contrast to the
previous case. Essentially, this occurs since the linear enhancement
can not completely compensate the exponential decay of evanescent
waves. However, this enhancement significantly improves transmission
characteristic of the free space slab. Therefore, the slabs of media
with extreme optical anisotropy can be successfully used for
subwavelength imaging purposes.

The proposed imaging technique can be viewed as an improvement of
the canalization regime \cite{canal,SWIWM,WMIR,layeredPRB} which
enables a transmission of images with subwavelength resolution from
the front- to back-interface of a slab formed by medium with extreme
optical anisotropy. Without any change in the structure, the imaging
device can operate with sources located at certain distance from the
slab, but the image in this case has to be detected inside the
medium.

The imaging devices are not perfect by any means since they
introduce certain distortions because their transmission coefficient
is not equal to unity for the entire spectrum of spatial harmonics.
However, the form of this distortion is known: it is given by
formula \r{Td2}. Thus, the distorted image captured inside of the
proposed device can be processed and the original distribution can
be restored. For this purpose, it is enough to expand the image into
spatial spectrum using a 2D Fourier transform, to divide it by the
transmission coefficient and to perform an inverse 2D Fourier
transform. Within the range of spatial frequencies corresponding to
large enough transmission, this procedure leads to perfect
restoration of the spectrum of the original image. The remainder of
the spatial information has to be filtered since the division by
small transmission coefficient leads to significant enhancement of
signal to noise ratio.

In conclusion, the phenomenon of evanescent waves enhancement
originated from resonant pumping of standing waves is observed
inside the slabs of media with extreme optical anisotropy. This
enables subwavelength imaging of objects placed at considerable
distances away from the slab. \phantom{\footnote{P.A.Belov
acknowledges financial support by EPSRC Advanced Research Fellowship
EP/E053025/1 and thanks George Palikaras for help in editing of the
manuscript.}}

\bibliography{internal}

\begin{thebibliography}{10}
\expandafter\ifx\csname natexlab\endcsname\relax\def\natexlab#1{#1}\fi
\expandafter\ifx\csname bibnamefont\endcsname\relax
  \def\bibnamefont#1{#1}\fi
\expandafter\ifx\csname bibfnamefont\endcsname\relax
  \def\bibfnamefont#1{#1}\fi
\expandafter\ifx\csname citenamefont\endcsname\relax
  \def\citenamefont#1{#1}\fi
\expandafter\ifx\csname url\endcsname\relax
  \def\url#1{\texttt{#1}}\fi
\expandafter\ifx\csname urlprefix\endcsname\relax\def\urlprefix{URL }\fi
\providecommand{\bibinfo}[2]{#2}
\providecommand{\eprint}[2][]{\url{#2}}

\bibitem[{\citenamefont{Fang et~al.}(2005)\citenamefont{Fang, Lee, Sun, and
  Zhang}}]{Silversub}
\bibinfo{author}{\bibfnamefont{N.}~\bibnamefont{Fang}},
  \bibinfo{author}{\bibfnamefont{H.}~\bibnamefont{Lee}},
  \bibinfo{author}{\bibfnamefont{C.}~\bibnamefont{Sun}}, \bibnamefont{and}
  \bibinfo{author}{\bibfnamefont{X.}~\bibnamefont{Zhang}},
  \bibinfo{journal}{Science} \textbf{\bibinfo{volume}{308}},
  \bibinfo{pages}{534} (\bibinfo{year}{2005}).

\bibitem[{\citenamefont{Pendry}(2000)}]{Pendrylens}
\bibinfo{author}{\bibfnamefont{J.}~\bibnamefont{Pendry}},
  \bibinfo{journal}{Phys. Rev. Lett.} \textbf{\bibinfo{volume}{85}},
  \bibinfo{pages}{3966} (\bibinfo{year}{2000}).

\bibitem[{\citenamefont{Jacob et~al.}(2006)\citenamefont{Jacob, Alekseyev, and
  Narimanov}}]{Narimanov}
\bibinfo{author}{\bibfnamefont{Z.}~\bibnamefont{Jacob}},
  \bibinfo{author}{\bibfnamefont{L.~V.} \bibnamefont{Alekseyev}},
  \bibnamefont{and}
  \bibinfo{author}{\bibfnamefont{E.}~\bibnamefont{Narimanov}},
  \bibinfo{journal}{Optics Express} \textbf{\bibinfo{volume}{14}},
  \bibinfo{pages}{8247} (\bibinfo{year}{2006}).

\bibitem[{\citenamefont{Shalaev}(2007)}]{Shalaev}
\bibinfo{author}{\bibfnamefont{V.~M.} \bibnamefont{Shalaev}},
  \bibinfo{journal}{Nature Photonics} \textbf{\bibinfo{volume}{1}},
  \bibinfo{pages}{41} (\bibinfo{year}{2007}).

\bibitem[{\citenamefont{Belov et~al.}(2003)\citenamefont{Belov, Marques,
  Maslovski, Nefedov, Silverinha, Simovski, and Tretyakov}}]{WMPRB}
\bibinfo{author}{\bibfnamefont{P.}~\bibnamefont{Belov}},
  \bibinfo{author}{\bibfnamefont{R.}~\bibnamefont{Marques}},
  \bibinfo{author}{\bibfnamefont{S.}~\bibnamefont{Maslovski}},
  \bibinfo{author}{\bibfnamefont{I.}~\bibnamefont{Nefedov}},
  \bibinfo{author}{\bibfnamefont{M.}~\bibnamefont{Silverinha}},
  \bibinfo{author}{\bibfnamefont{C.}~\bibnamefont{Simovski}}, \bibnamefont{and}
  \bibinfo{author}{\bibfnamefont{S.}~\bibnamefont{Tretyakov}},
  \bibinfo{journal}{Phys. Rev. B} \textbf{\bibinfo{volume}{67}},
  \bibinfo{pages}{113103} (\bibinfo{year}{2003}).

\bibitem[{\citenamefont{Silveirinha}(2006)}]{Silv_Nonlocalrods}
\bibinfo{author}{\bibfnamefont{M.}~\bibnamefont{Silveirinha}},
  \bibinfo{journal}{Phys. Rev. E} \textbf{\bibinfo{volume}{73}},
  \bibinfo{pages}{046612} (\bibinfo{year}{2006}).

\bibitem[{\citenamefont{Belov and Hao}(2006)}]{layeredPRB}
\bibinfo{author}{\bibfnamefont{P.}~\bibnamefont{Belov}} \bibnamefont{and}
  \bibinfo{author}{\bibfnamefont{Y.}~\bibnamefont{Hao}},
  \bibinfo{journal}{Phys. Rev. B} \textbf{\bibinfo{volume}{73}},
  \bibinfo{pages}{113110} (\bibinfo{year}{2006}).

\bibitem[{\citenamefont{Belov et~al.}(2005)\citenamefont{Belov, Simovski, and
  Ikonen}}]{canal}
\bibinfo{author}{\bibfnamefont{P.~A.} \bibnamefont{Belov}},
  \bibinfo{author}{\bibfnamefont{C.~R.} \bibnamefont{Simovski}},
  \bibnamefont{and} \bibinfo{author}{\bibfnamefont{P.}~\bibnamefont{Ikonen}},
  \bibinfo{journal}{Phys.~Rev.~B.} \textbf{\bibinfo{volume}{71}},
  \bibinfo{pages}{193105} (\bibinfo{year}{2005}).

\bibitem[{\citenamefont{Belov et~al.}(2006)\citenamefont{Belov, Hao, and
  Sudhakaran}}]{SWIWM}
\bibinfo{author}{\bibfnamefont{P.~A.} \bibnamefont{Belov}},
  \bibinfo{author}{\bibfnamefont{Y.}~\bibnamefont{Hao}}, \bibnamefont{and}
  \bibinfo{author}{\bibfnamefont{S.}~\bibnamefont{Sudhakaran}},
  \bibinfo{journal}{Phys.~Rev.~B} \textbf{\bibinfo{volume}{73}},
  \bibinfo{pages}{033108} (\bibinfo{year}{2006}).

\bibitem[{\citenamefont{Silveirinha et~al.}(2007)\citenamefont{Silveirinha,
  Belov, and Simovski}}]{WMIR}
\bibinfo{author}{\bibfnamefont{M.}~\bibnamefont{Silveirinha}},
  \bibinfo{author}{\bibfnamefont{P.}~\bibnamefont{Belov}}, \bibnamefont{and}
  \bibinfo{author}{\bibfnamefont{C.}~\bibnamefont{Simovski}},
  \bibinfo{journal}{Phys. Rev. B} \textbf{\bibinfo{volume}{75}},
  \bibinfo{pages}{035108} (\bibinfo{year}{2007}).

\end{thebibliography}
\end{document}